\documentclass[11pt]{article}

\usepackage[dvips]{graphicx}

\def\bTheta{{\bf \Theta}}
\def\br{{\bf r}}
\def\bk{{\bf k}}
\def\bE{{\bf E}}
\def\bB{{\bf B}}
\def\bJ{{\bf J}}
\newcommand{\half}{{\scriptstyle{\frac{1}{2}}}}
\newcommand{\wbn}{{\widehat{\bf n}}}
\newcommand{\wbk}{{\widehat{\bf k}}}
\newcommand{\p}{{\partial}}
\def\parag{\hfil\break} 
\def\kikezd{\parag\underbar}

\begin{document}

\title{Anomalous Hall Effect in non-commutative mechanics}

\author{P.~A.~Horv\'athy
\footnote{e-mail: horvathy@lmpt.univ-tours.fr.} 
 \\ 
 Laboratoire de Math\'ematiques et de Physique Th\'eorique 
 \\
 Universit\'e de Tours\\  Parc de Grandmont\\ 
 F-37200 TOURS (France)
}

\maketitle

\begin{abstract}
The anomalous velocity term in the semiclassical model of a Bloch electron deviates the trajectory from
the conventional one. 
When the Berry curvature (alias noncommutative parameter) 
is a  mo\-nopole in momentum space, as found recently in
some ferromagnetic crystals while observing
the anomalous Hall effect, we get a transverse shift,
similar to that in the optical Hall effect.
\end{abstract}
\vspace{5mm}\noindent
\texttt{cond-mat/0606472}

\section{Introduction}
       
The Anomalous Hall Effect (AHE), characterized
by the absence of a magnetic field, is observed in some
ferromagnetic crystals. While this
has been well established experimentally,
its explanation is still controversial.
One, put forward by Karplus and Luttinger \cite{Lutti}
 fifty years ago, suggests that the effect is due to
an anomalous current.

 Many years later,
 it  has been argued \cite{Niu} that the semiclassical dynamics of a Bloch electron in 
 a crystal should involve a Berry curvature term,
$\bTheta$.
In the $n{}^{th}$ band the equations of motion read, in an electromagnetic field, 
\begin{eqnarray}
\dot{\bf r}&=\displaystyle\frac{\partial\epsilon_n(\bk)}
{\partial\bk}-\dot\bk\times{\bTheta},
\label{velrel}
\\[6pt]
\dot{\bk}&=e{\bf E}+e\dot{{\bf r}}\times{\bf B}({\bf r}),
\label{Lorentz}
\end{eqnarray}
where ${\bf r}=(x^i)$ and $\bk=(k_j)$ denote the electron's 
intracell position and quasimomentum,
respectively; $\epsilon_n(\bk)$ is the band energy.
The relation (\ref{velrel}) exhibits 
an \textit{anomalous velocity term},
$\dot\bk\times{\bTheta}$, which is
the mechanical counterpart of the anomalous current.

The model is distinguished by the non-commutativity
of the position coordinates~: 
in the absence of a magnetic field, 
 $\{x^i,x^j\}=\epsilon^{ijn}\Theta_n$ \cite{BeMo,DHHMS}.
In the free noncommutative model in
$3$ space dimensions, $\bTheta$ can only be
momentum-dependent such that $\p_{k_i}
\Theta^i=0$ \cite{BeMo}. 

A remarkable discovery concerns
the AHE in the metallic ferromagnet SrRuO${}_3$.
Fang et al. \cite{Fang} found in fact that the 
experimental data are consistent with $\bTheta$ 
taking the form of a \textit{monopole in momentum space},
\begin{equation}
{\bTheta}=\theta\frac{\bk}{k^3},
\label{kmonop}
\end{equation}
$k\neq0$. (\ref{kmonop}) is, furthermore, the
only possibility consistent with rotational symmetry  
\cite{BeMo}.

Here we propose to study the AHE in the
semiclassical framework (as advocated in \cite{AHE}),
with non-commutative parameter (\ref{kmonop}).
For $\bB=0$ and a constant electric field,
 $\bE=\,$const. and assuming a parabolic profile 
$\epsilon_n(\bk)=\bk^2/2$, eqn. (\ref{Lorentz}),  $\dot{\bk}=e\bE$, is integrated as 
$ 
\bk(t)=e\bE\,t+\bk_0.
$ 
The velocity relation (\ref{velrel}) becomes in turn
\begin{eqnarray}
\dot{\bf r}=\bk_0+e\bE t
+\frac{e\theta Ek_0}{k^3}\,\wbn,
\label{AHEvelrel}
\end{eqnarray}
where  $\wbn=\wbk_0\times\widehat{\bE}$
[``hats'' denote vectors
normalized to unit length]. 
The component of $\bk_0$ parallel to $\bE$ has no interest;
we can assume therefore that $\bk_0$ is perpendicular
to the electric field. Writing 
$\br(t)=x(t)\wbk_0+y(t)\widehat{\bE}+z(t)\wbn$,
eqn. (\ref{AHEvelrel}) yields that the component parallel to
$\bk_0$ moves uniformly, $x(t)=k_0t$, and its component parallel
to the electric field is uniformly accelerating, $y(t)=\half eEt^2$. (Our choices correspond
to chosing time so that the turning point is at $t=0$.)
However, owing to the anomalous term in (\ref{velrel}),
the particle is  also deviated perpendicularly to $\bk_0$ and $\bE$, 
namely by
\begin{equation} 
z(t)=\frac{\theta}{k_0}\frac{eEt}{\sqrt{k_0^2+e^2E^2t^2}}.
\end{equation}
It follows that the trajectory leaves its initial the plane
and suffers, between $t=-\infty$ to $t=\infty$,  a 
\textit{finite}
\textit{transverse shift}, namely
\begin{equation}
\Delta z=\frac{2\theta}{k_0}.
\label{shift}
\end{equation}
\goodbreak
Most contribution to the shift comes when the momentum is small, i.e., ``near the $\bk$-monopole.''

$\theta$ becomes a half-integer upon quantization,
$\theta=N/2$, and hence (\ref{shift}) is indeed $N/k_0$.
The  constant $k_0\neq0$, the minimal possible value of
momentum, plays the role of an impact parameter.
Let us observe that while (\ref{shift}) does not
depend on the field $\bE$ or the electric charge $e$,
the limit $eE\to0$ is singular. For $eE=0$, the motion
is uniform along a straight line.
\begin{figure}
\includegraphics[scale=.9]{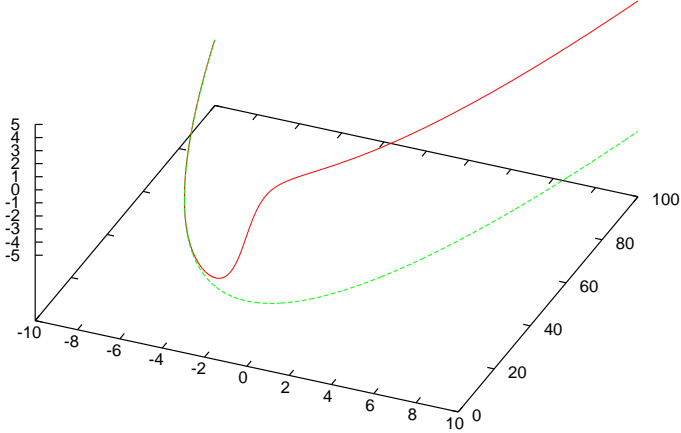}
\label{AHEplot}
\caption{\it The anomalous velocity term deviates the trajectory form the plane.}
\end{figure}

The transverse shift is reminiscent of the recently discovered optical Hall effect
\cite{OHE} and can also be derived, just like in the optical case, 
 using the  angular momentum.
The free expression \cite{BeMo},
$ 
\bJ=\br\times\bk-\theta\,\wbk,
$ 
is plainly broken by the electric field
 to its component parallel to $\bE$,
\begin{equation}
J=J_y=z(t)k_0-\theta\,\frac{eEt}{\sqrt{k_0^2+e^2E^2t^2}},
\end{equation}
whose conservation yields once again the shift (\ref{shift}).

 How can the same argument work for a Bloch electron
and for light~? 
The answer relies, for both problems, on having
the same ``$\bk$-monopole'' contribution, $-\theta\wbk$
in the angular momentum.
\goodbreak
 
Our model is plainly not realistic: what we described is, rather, the
deviation of a freely falling non-commutative particle  from the classical
parabola found by Galileo. Particles in a metal are not free,
though, and their uniform acceleration
in the direction of $\bE$ should be damped by some
mechanism. It is nevertheless remarkable that we obtain 
qualitative information from such a toy model.

\kikezd{Note added}. 
I am indebted to Dr. S. Murakami for calling my attention to
similar work in the context of 
the spin Hall effect in semiconductors \cite{SpinHall}. 
I would also like to thank Dr. Y. Kats for informing me
about the experimental status of the AHE, see \cite{Kats}.


\end{document}